\documentclass{iopart}
\newcommand{\tnote}[1]{}
\usepackage{braket}
\usepackage{graphicx}

\newcommand{\eqref}[1]{(\ref{#1})}

\renewcommand{\-}{\nonumber\\}

\newcommand{\authornote}[1]{}
\newcommand{\vect}[1]{\vec{#1}}
\renewcommand{\Im}{{\rm Im \;}}
\newcommand{\expec}[1]{\ensuremath{\left\langle #1 \right\rangle}}

\begin{document}
\title[Spin-boson models for quantum decoherence]{Spin-boson models for quantum decoherence of electronic excitations of biomolecules and quantum dots in a solvent}
\author{Joel Gilmore and Ross H. McKenzie}
\address{Department of Physics, University of Queensland, Brisbane, Qld 4072, Australia}
\date{\today}
\begin{abstract}
We give a theoretical treatment of the interaction of electronic excitations (excitons) in biomolecules and quantum dots with the surrounding polar solvent.  Significant quantum decoherence occurs due to the interaction of the electric dipole moment of the solute with the fluctuating electric dipole moments of the individual molecules in the solvent.  We introduce spin boson models which could be used to describe the effects of decoherence on the quantum dynamics of biomolecules which undergo light-induced conformational change and on biomolecules or quantum dots which are coupled by F\"orster resonant energy transfer.
\end{abstract} 

\maketitle

\section{Introduction}

Biomolecular systems contain a diverse range of optically active molecules that are crucial to biological function \cite{Helms02}. Important examples include retinal in rhodopsin which plays a role in vision, porphyrins and chlorophyls which play roles in photosynthesis \cite{Hu97PhysicsToday}, photoactive yellow protein, blue copper proteins, and the green fluorescent protein (GFP) \cite{Sullivan99} which has become a powerful tool in molecular and cell biology.  The different chromophores couple very differently to their environment which consists of the surrounding proteins and the solvent (water or aqueous electrolytes) leading to very different quantum dynamics \cite{DecayNote}.  Related questions arise concerning the spatial extent of the quantum coherence of excitations in spatially extended structures such as $\alpha$-helices and $\beta$-sheets in proteins and rings of porphyrin molecules in light harvesting complexes \cite{Damjanovic02}. Thus, it is important to understand the coupling of electronic excitations in chromophores to the environment. This is not just of interest for understanding biological function, but also because these biomolecular systems could potentially be model systems to understand quantum decoherence at the nanoscale. This is because they have a complexity that cannot be fabricated even with the most advanced nanotechnology, but they can be modified and tuned using sophisticated biochemical techniques \cite{Niemeyer00}. We also note that there are potential applications in biosensors with biomolecules coupling to quantum dots via resonant energy transfer \cite{Medintz03}, and with biomolecules acting as  optically controllable elements in molecular electronics \cite{Birge99,Cinelli01} or as qubits in a quantum computer \cite{Lovett03PRB}.

Understanding the quantum dynamics of a system that is strongly coupled to its environment is a highly non-trivial, and sometimes controversial, problem. Furthermore, given the complexity and diversity of biomolecular systems it is desirable to find tractable model Hamiltonians which capture the essential physics and are amenable to rigorous analysis. The past few decades have seen monumental advances in combined quantum mechanical and molecular dynamics simulations, allowing the study of biomolecular systems of realistic size. See for example, the recent study of the quantum dynamics of retinal in the rhodopsin environment \cite{Hayashi03}.  However, in spite of this progress it is unrealistic that in the next decade it will be possible to give a full treatment of the quantum dynamics including the quantum many-body effects associated with the strong interaction between the chromophore and its environment. Furthermore, simple models amenable analytical treatment can give insight into the essential physics involved with qualitative differences, especially when model parameters are extracted from quantum chemistry and molecular dynamics.''

We first consider the problem of the interaction of an electronic excitation (exciton) in a biomolecule (or a quantum dot) with the surrounding polar solvent. We look for the simplest possible model which can still describe the relevant physics, which allows us to determine the important parameters and perform analytical calculations. We show that this system can be described by an independent boson model where the environment is modelled as an infinite collection of harmonic oscillators with ohmic response.  In the model we derive, the strength of the coupling to the environment is determined by the frequency dependent dielectric function of the polar solvent and by the difference between the electric dipole moments of the molecule in the ground and excited states. The relaxation rate of the polar molecules within the solvent determines the cutoff frequency for the ohmic response.  

We then introduce minimal models which can describe the effect of the solvent on the quantum dynamics associated with conformational changes at conical intersections \cite{Yarkony96, Molnar00, Hahn00}, and on excitations which are coupled via a mechanism such as resonance energy transfer. The latter is important in photosynthesis \cite{Hu97PhysicsToday}, is the basis of Fluoresence Resonance Energy Transfer (FRET) spectroscopy which is used to determine distances between chromophores in biomolecules \cite{Wu94,Ha01}, and is the basis of new biosensors \cite{Medintz03}.  We have shown that each may be modelled by spin-boson models \cite{Weiss99,Leggett87} which exhibit rich many-body physics, and have been used to model systems such as the coupling between electron transfer and protein motion or a solvent \cite{Weiss99,Xu94ChemPhys,Renger02,Fleming96}.  The general spin-boson Hamiltonian is 
\begin{equation}\label{eq:spin boson defn}
H = \frac{1}{2} \epsilon \sigma_z - \frac{1}{2} \Delta \sigma_x + \sum_\alpha \omega_\alpha a_\alpha^\dagger a_\alpha + \frac{1}{2}\sigma_z \sum_\alpha C_\alpha (a_\alpha+a_\alpha^\dagger),
\end{equation}
where $\sigma_z,\sigma_x$ are Pauli matrices describing the TLS, $\epsilon$ is the energy separation of the two levels, $\Delta$ is the tunneling matrix element between the two levels, and the bath is modelled by boson operators $a_\alpha,a_\alpha^\dagger$ with frequencies $\omega_\alpha$ and couplings $C_\alpha$ to the TLS.  The independent boson model is \eqref{eq:spin boson defn} with $\Delta=0$.  This means that, unlike the spin-boson model, there are no transitions between the two quantum states. Only one environmental parameter, however, contributes to the system dynamics, the \emph{spectral density} \cite{Leggett87,Weiss99}
\begin{equation}\label{eq:J(omega) defn}
J(\omega)= 4\pi \sum_\alpha C_\alpha^2 \delta(\omega-\omega_\alpha).
\end{equation}

Leggett \emph{et al}. \cite{Leggett87} defined a dimensionless coupling constant $\alpha$ which determines the extent of the quantum coherence. We show that for realistic biomolecules solvated in water this parameter will typically have values of order unity, which would prevent coherent Bloch oscillations for degenerate electronic levels. Such large values are in stark contrast to the Josephson junction qubits for which the effect of the Johnson noise in the electronic circuit is described by a spin boson model with values of $\alpha$ that are typically many orders of magnitude smaller \cite{Makhlin01}. 

\section{Independent boson model for the interaction of a single chromophore with a solvent.} 

We now show how an independent boson model can describe an electronic excitation in a biomolecule or quantum dot coupled to a solvent bath.  We model the chromophore as a TLS with only a single active transition, of energy $\epsilon$.  We assume the solute molecule has a permanent dipole moment in both the ground and excited state, given by $\vect{\mu}_g$ and $\vect{\mu}_e$ respectively.  Describing the electronic ground and excited states by fermion creation operators $c_g^\dagger$ and $c_e^\dagger$ respectively, the solute part of the total Hamiltonian operator is
\begin{equation}
H_{TLS} = \frac{1}{2} \epsilon( c_e^\dagger c_e - c_g^\dagger c_g ).
\end{equation}

\begin{figure}[tbp]
\begin{center}
\includegraphics[width=6cm]{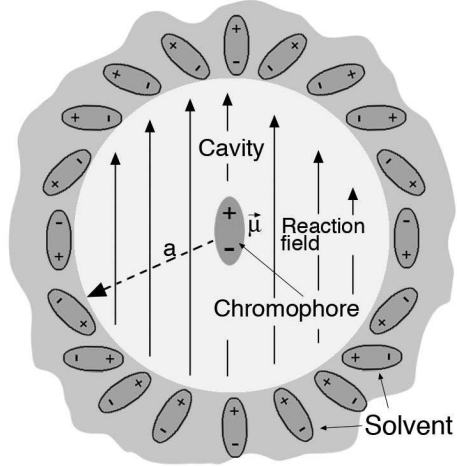}
\caption{Onsager model of solvation.  The chromophore is treated as a point dipole $\vect{\mu}$ inside an empty cavity of radius $a$, which is surrounded by a polar solvent. The dipole polarises the solvent which in turn creates a an electric ``reaction'' field which acts back on the dipole, stabilising the solvated system.}
\label{fig:onsager}
\end{center}
\end{figure}

In the simplest picture, we can describe the solvation process by the well-known Onsager model \cite{Onsager36,Nemkovich91}.  This is a type of \emph{continuum model} where the solvent is treated as a homogeneous dielectric, and has been used for both analytic and computational studies of solvation \cite{Simonson01, Hofinger01}.  The solute is treated as a point dipole which is surrounded by a spherical cage of polar solvent molecules with Onsager radius $a$ (Figure \ref{fig:onsager}), which is typically the size of the solute molecule \cite{Onsager36,Nemkovich91,Tomasi94}.  The cavity is assumed to be a vacuum, i.e., it has a dielectric constant $\epsilon_r = 1$.   The hydration shell of water molecules who have broken hydrogen bonds with their neighbours, and are now bonded to the biomolecule, will serve to increase the effective cavity radius $a$. Further details (e.g., surrounding proteins) can be treated by adding shells with different dielectric constants around this cavity; this will be considered in a later work.

The central dipole polarises the cage, which in turn produces an electric field inside the cavity, called the \emph{reaction field} $\vect{R}(t)$.  For the case of a uniform spherical cavity the reaction field is constant everywhere inside \cite{Bottcher73}. This field acts back on the dipole, with interaction energy $E = - \vect{\mu(t)} \cdot \vect{R(t)}$, typically lowering the total energy and hence forming a stable structure.  Although the Onsager model does not include the microscopic details of the system, it does capture the essential physics of the solvation process \cite{Lakowicz99a} and provided due care is taken to be aware of important microscopic effects (such as charge transfer processes) it should still provide valuable physical insight.

Prior to the electronic transition, we assume the system to be in thermal equilibrium, and the reaction field will be parallel to the dipole $\vect{\mu}_g$.  For simplicity, we assume that the direction of the dipole moment of the solute molecule in the ground and excited states points in the same direction. Defining $\Delta\mu = \mu_e-\mu_g$, $\sigma_z \equiv c_e^\dagger c_e - c_g^\dagger c_g$, and identifying $c_e^\dagger c_e + c_g^\dagger c_g$ with the $2\times 2$ identity, \tnote{so that}
\begin{eqnarray*}
H_{solute} = \frac{1}{2} \epsilon \sigma_z,  & \qquad H_{int} &= \frac{1}{2} \sigma_z (\Delta {\mu}) R  + \frac{1}{2} (\mu_e+\mu_g) R.
\end{eqnarray*}

For a quantum mechanical model, we must quantize the reaction field. In the Heisenberg picture, we express the reaction field $R(t)$ in its Fourier transform modes,
\begin{equation}
R(t) = \sum_\alpha e_\alpha \left[ a_\alpha e^{- i\omega_\alpha t}+ a_\alpha^\dagger e^{i\omega_\alpha t} \right],
\end{equation}
and quantise the coefficients so that $a_\alpha$ and $a_\alpha^\dagger$ obey the boson commutation relations $[a_\alpha,a_\beta^\dagger]=\delta_{\alpha,\beta}$.

We are therefore modelling the environment (solvent) as a bath of independent harmonic oscillators. The energy stored in the solvent cage is now expressed as $H_{solvent} = \sum_\alpha \omega a_\alpha^\dagger a_\alpha$ and the full Hamiltonian for a single biomolecule coupled to the environment can then be written as
\begin{eqnarray}\label{eq:spinboson final}
H &= \frac{1}{2} \epsilon \sigma_z + \sum_\alpha \omega_\alpha a_\alpha^\dagger a_\alpha + \-
& \quad \sigma_z \sum_\alpha M_\alpha (a_\alpha^\dagger + a_\alpha) + \sum_\alpha \widetilde{M}_\alpha (a_\alpha^\dagger+a_\alpha),
\end{eqnarray}
where we define couplings \mbox{$M_\alpha = \frac{1}{2} e_\alpha \Delta \mu$} and {$\widetilde{M}_\alpha = \frac{1}{2} e_\alpha (\mu_e+\mu_g)$}.  

This is the independent boson model \cite{Mahan90a,ExtraTermNote}. We have not included a $\sigma_x$ term, so that there can be no transitions between the two energy levels of the chromophore and the chromophore energy $\epsilon \expec{\sigma_z}$ is a constant. Two conditions are needed to justify this.  First, we consider only times after the initial excitation of the system (e.g., by a laser pulse) but much shorter than the radiative lifetime of the chromophore (typically 1-10ns \cite{vanHolde98a}).  This means we do not include coupling of the TLS to the quantum and thermal fluctuations of the electric field which can induce optical transitions. At energies of the order of eV these fluctuations are much smaller than the dielectic fluctuations associated with the solvent.  Secondly, for many electronic excitations we have assumed that the energy gap $\epsilon$ (typically of order of 1eV) is significantly larger than the tunneling matrix element, and so its effects can be ignored on the timescales of interest. We defer discussion of it until below when we consider $\epsilon$ becoming small near a conformational transition.

The independent boson model can be used to investigate decoherence of systems coupled to a bath.  Consider an initially coherent excitation, i.e., a coherent superposition of the ground and excited states of the chromophore.  In the reduced density matrix of the chromophore, the population of each level is given by the diagonal elements, while the coherence is given by the off-diagonal elements. Thus the initial reduced density matrix of the chromophore has non-zero off-diagonal terms. Reference \cite{Reina02} shows that while the populations remain constant (as expected) the coherence decays exponentially, as given by Equations (18) and (19) and Figure 7 of \cite{Reina02}.  Hence, after long times the reduced density matrix has no diagonal terms and so describes the chromophore in a mixed state \cite{Costi03}.

The above model therefore describes the decoherence of an initially coherent excitation of a chromophore due to its environment (water, proteins),  as well as spectral features such as the shifting and broadening of the absorption peak, as will be described below. 

\section{Derivation of the spectral density function} 

To complete the model we must specify the environmental coupling, i.e., the spectral function $J(\omega)$ from \eqref{eq:J(omega) defn}.  Following an approach similar to Caldeira and Leggett \cite{Caldeira83} who considered a Josephson junction coupled to an electronic circuit, we relate $J(\omega)$ to the zero-temperature fluctuations in the uncoupled environment \cite{Fleming96}.  Noting that with no solute-solvent interaction $\expec{R(t)}=0$, we now examine the reaction field fluctuation correlation function $S(t)$ \cite{Rickayzen80} defined as
\begin{equation}\label{eq:tcf}
S(t) = i\expec{R(t)R(0)}\theta(t) \equiv i\Braket{0|e^{iHt} R e^{-iHt} R|0} \theta(t),
\end{equation}
where $\ket{0}$ is the ground state of the solvent harmonic oscillators, and $H\equiv H_{solvent} = \sum_\alpha \omega_\alpha a_\alpha^\dagger a_\alpha$.  We shall see that it is the imaginary part of the Fourier transform of $S(t)$, which we denote $\mathcal{E}(\omega)$, that is relevant. We can write this as a sum over energy eigenstates \cite{Rickayzen80},
\begin{equation}
\Im \mathcal{E}(\omega) = \pi {\sum}_n \delta(\omega-E_n) \left|\Braket{0|R|n}\right|^2.
\end{equation}
Expanding $R$ into its normal modes again, we see that all terms $\Braket{0|R|n}$ vanish except for when a single oscillator is singly occupied.  These states have energy $E_n = \omega_\alpha$, and $\Braket{0|R|\alpha}=e_\alpha$.  Therefore, we see $\Im \mathcal{E}(\omega) = \pi \sum_\alpha e_\alpha^2 \delta(\omega - \omega_\alpha)$, and using \eqref{eq:J(omega) defn}, we find
\begin{equation}\label{eq:spectral fluctuation connection}
J(\omega) = (\Delta \mu)^2 \Im \mathcal{E}(\omega).
\end{equation}
\tnote{We must still, however, calculate the fluctuations in the reaction field.  In the Onsager model \cite{Onsager36,Hsu97}, the reaction field $R(t)$ and the central dipole $\mu(t)$ are related by a linear response function $\chi(t-t')$}
To calculate the reaction field fluctuations, we note that in the Onsager model \cite{Onsager36,Hsu97}, $R(t)$ and the central dipole $\mu(t)$ are related by a linear response function $\chi(t-t')$ \cite{Kubo66}, such that in Fourier space $R(\omega) = \chi(\omega)\mu(\omega)$.  The static limit, where the solvent cage adjusts instantaneously, so that $R(t)= \chi_s \mu(t)$, is well known \cite{Bottcher73,Onsager36}. In reality, the solvent cage will lag behind a changing dipole due to dielectric friction \cite{Nee70,Maroncelli97},and $\chi(\omega)$ is given \cite{Hsu97} by 
\begin{equation}\label{eq:susceptibility}
\chi(\omega) =  \frac{1}{4\pi\epsilon_0 a^3} \frac{2(\epsilon(\omega)-1)}{2\epsilon(\omega)+1}.
\end{equation}
where $\epsilon(\omega)$ is the frequency dependent dielectric constant of the solvent. We use the Debye formula \cite{Song93,DebyeNote}
\begin{equation}\label{eq:debye model}
\epsilon(\omega) = \epsilon_\infty + \frac{\epsilon_s-\epsilon_\infty}{1-i\omega\tau_D},
\end{equation}
where $\epsilon_\infty$ and $\epsilon_s$ are the high and low (static) frequency limits respectively, and $\tau_D$ is the Debye relaxation time, the bulk reorientational relaxation time of the solvent dipoles \cite{Kindt96,Barthel91}.  For water, these are given by $\epsilon_s=78.3$, $\epsilon_\infty=4.21$ and $\tau_D = 8.2\rm{ps}$ \cite{Afsar78}. 

We then apply the quantum fluctuation-dissipation relation \cite{Chaikin95a, Kubo66}, for the imaginary part of $\chi(\omega)$
\begin{equation}
\chi''(\omega) = -(i/2) \left( 1 - e^{-\beta\omega} \right) \mathcal{E}(\omega),
\end{equation}
which reduces at zero temperature to $\mathcal{E}(\omega) = 2 i  \chi''(\omega)$. Note that this limit is well defined, even for $\omega=0$.  Using \eqref{eq:debye model} and \eqref{eq:spectral fluctuation connection}, the spectral density is then given by
\begin{equation}\label{eq:J(omega) final}
J(\omega) =\frac{(\Delta\mu)^2}{2\pi\epsilon_0 a^3} \frac{6(\epsilon_s-\epsilon_\infty)}{(2\epsilon_s+1)(2\epsilon_\infty+1)} \frac{\omega\tau_E}{\omega^2\tau_E^2+1},
\end{equation}
where $\tau_E = \frac{2\epsilon_\infty+1}{2\epsilon_s+1} \tau_D$. 

This form for $J(\omega)$, which has a specific microscopic basis, is an important new result and could be used as an input into more phenomenological models such as the quantum Brownian model \cite{Mukamel95a}.   In specific situations the magnitude of the $J(\omega)$ introduced here should also be compared to the $J(\omega)$ due to slow protein motion, as introduced in \cite{Xu94ChemPhys} and \cite{Renger02}.

We have thus shown that solute-solvent interaction in the Onsager picture can be modeled by a independent boson model with spectral density \eqref{eq:J(omega) final}. We note that $J(\omega)$ has an implicit high frequency cut-off at $\omega_c= \frac{1}{\tau_E}$, related to the finite relaxation time of the solvent dipoles.  Below the cut-off, $J(\omega)$ is approximately linear: $J(\omega) = \eta\omega$, where
\begin{equation}\label{eq:eta}
\eta = \frac{(\Delta\mu)^2}{4\pi\epsilon_0 a^3} \frac{6(\epsilon_s-\epsilon_\infty)}{(2\epsilon_s+1)^2} \tau_D .
\end{equation}
Thus, our model falls into the class of models defined in Ref. \cite{Leggett87} which below some high frequency cutoff have Ohmic dissipation.  $\eta$ should correspond to the classical frictional coefficient.  Indeed, $\eta$ is identical to the  Nee-Zwanzig prediction of dielectric friction \cite{Nee70,Maroncelli97,Papazyan95}.

\section{Derivation of the optical absorption lineshape}

We can also use this model to derive an expression for the absorption lineshape $\alpha(\omega)$ of the solute molecule in the presence of the solvent. In the gas phase, i.e., without the solvent, we would expect a sharp peak in the absorption spectrum at $\omega = \epsilon$.  In the presence of the solvent, however, this peak is shifted and broadened \cite{Bayliss54,Nemkovich91,Myers98} by the solute-solvent dipole-dipole interactions.

We use the results obtained for the \emph{independent boson model} \cite{Mahan90a}.  The resulting spectrum is
\begin{eqnarray}
\alpha(\omega) &= \frac{A}{\omega} \rm{Re} \int_{-\infty}^{\infty} dt \exp\left\{ it\left[\omega-\epsilon+\widetilde\Delta-g(t)\right]\right\},
\end{eqnarray}
where $A$ is a normalisation factor (which we are not concerned with), and $\widetilde\Delta = 4\sum_\alpha \frac{M_\alpha \widetilde{M}_\alpha}{\omega_\alpha}$.  $g(t)$ is the factor responsible for broadening the lineshape \cite{Mukamel95a}.  We express it \cite{Leggett87,Mukamel95a} using the spectral density \eqref{eq:J(omega) defn}:
\begin{equation*}
g(t) = \int d\omega \frac{J(\omega)}{\pi\omega^2} \left[ \coth\left(\frac{\beta\omega}{2}\right)\left(1-\cos\left(\omega t\right)\right)+i\sin\left(\omega t\right) \right].
\end{equation*}
Since $J(\omega) \approx 0$ for $\omega \gg \omega_c$, we can restrict the integral to $\omega<\omega_c$, and note that for water at room temperature (300 K, $\tau_D \approx 8$ ps \cite{Afsar78}) $\beta\omega < \beta\omega_c \ll 1$. We also make the short-time approximation \cite{Mukamel95a} and Taylor expand the trigonometric functions to give $g(t)=\frac{1}{2} k_B T \lambda_0 t^2 + \frac{i}{2} \lambda_0 t$, where we have defined the reorganisation energy \mbox{$\lambda_0 = \frac{2}{\pi} \int \frac{J(\omega)}{\omega} d\omega$}. We note also that because $\widetilde{M}_\alpha = \frac{\mu_e+\mu_g}{\Delta \mu}M$, we also have $\widetilde\Delta = \frac{\mu_e+\mu_g}{2(\Delta \mu)} \lambda_0$. Performing the Gaussian integral, we obtain the absorption spectrum
\begin{equation*}
\alpha(\omega) = \frac{A}{\omega\sqrt{k_B T \lambda_0}} \exp \left\{ - \left[\frac{\omega-\left(\epsilon-({\mu_g}/{\Delta\mu)}\lambda_0\right)}{\sqrt{2k_B T \lambda_0}} \right]^2 \right\},
\end{equation*}
where we have renormalised $A$, and
\begin{equation}
\lambda_0 = \frac{(\Delta\mu)^2}{4\pi\epsilon_0 a^3} \frac{6(\epsilon_s-\epsilon_\infty)}{(2\epsilon_s+1)(2\epsilon_\infty+1)}.
\end{equation}
If we set $\epsilon_\infty=1$ then the resulting lineshape is precisely that of \cite{Nemkovich91}, where the absorption spectrum was obtained classically by assuming a Boltzmann distribution of the environment.  Thus, our quantum model gives the correct classical limit.

\section{Spin-boson models of conformational change through conical intersections} 

This model of solvent interactions can be extended to other situations.  A number of important photochemical reactions in biomolecules involve a conformational change that initiates a signal.  Examples include the \emph{cis}/\emph{trans} isomerisation of retinal \cite{Molnar00} and the photoactive yellow protein (PYP) \cite{Yoda01}.  The former drives the proton pumping (in rhodopsin) against an external pH gradient that leads to the signal responsible for vision.  The latter causes a conformational transition of the surrounding PYP protein.  Only recently has it been shown \cite{Molnar00,BenNun02} that these isomerisations occur through a ``conical intersection'' \cite{Yarkony96}, between the potential energy surfaces of the ground and excited states.  

\begin{figure}[tbp]
\begin{center}
\includegraphics[width=12cm]{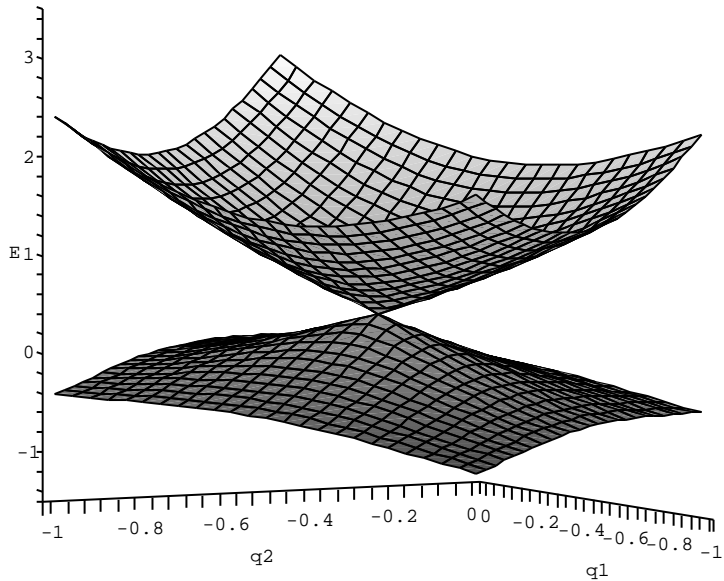}
\caption{A plot of two energy surfaces intersecting in a conical intersection.  The vertical axis is energy, and the horizontal axes are $q_1$ and $q_2$ representing the position co-ordinates of two classical harmonic oscillators.  Scales are in arbitrary units.}
\label{fig:CI}
\end{center}
\end{figure}

A minimal model for this requires two nuclear co-ordinates (degrees of freedom) \cite{BenNun02}, which form a two-dimensional simple harmonic oscillator, corresponding to the classical ``cone shaped'' potential surfaces \cite{Yarkony96}. The simplest model, with classical nuclear co-ordinates, which can describe a conical  intersection is
\begin{equation}
\fl H = \epsilon \sigma_z + \Delta \sigma_x + \frac{1}{2m_a} (p_a^2 + m_a^2 \omega_a^2 q_a^2) + \frac{1}{2m_b} (p_b^2+ m_b^2 \omega_b^2 q_b^2) + g_a \sigma_z q_a + g_b \sigma_x q_b,
\end{equation}
where $q_{a,b}$ and $p_{a,b}$ represent the position and momentum respectively of the two classical oscillators, $\omega_{a,b}$ and $m_{a,b}$ are the frequencies and masses, respectively, of the oscillators, and $g_{a,b}$ is the coupling of each oscillator to the chromophore.  This model includes a coupling term $\Delta \sigma_x$ between the two states, where $\Delta$ is the matrix element responsible for the non-adiabatic mixing of the ground and excited states \cite{Yarkony96}.  The spectrum is
\begin{equation}
E_{\pm} = \pm \sqrt{ \left(\epsilon+g_a q_a\right)^2 + \left(\Delta + g_b q_b\right)^2} +  \frac{1}{2} (\omega_a^2 p_a^2 +q_a^2 + \omega_b^2 p_b^2 + q_b^2)
\end{equation}
which clearly has a conical intersection when the first term vanishes (Figure \ref{fig:CI}).  We note that for $\epsilon=0$ this corresponds to the $E \otimes \epsilon$ Jahn-Teller model \cite{Hines04}.  Hahn and Stock \cite{Hahn00} recently studied such a minimal model for retinal in rhodopsin but did not include the effect of the solvent, which can play a pronounced role on the topology of the conical intersection \cite{Burghardt04}.  The generalisation of this model to include the solvent  and quantum motion of the nuclear coordinates is
\begin{eqnarray}
\fl H = \frac{1}{2} \epsilon \sigma_z + \Delta \sigma_x  + \sigma_z \sum_\alpha M_\alpha (a_\alpha^\dagger + a_\alpha) +  \sum_\alpha \omega_\alpha a_\alpha^\dagger a_\alpha + \\
\omega_a a^\dagger a + \omega_b b^\dagger b \; + g_a \sqrt{\frac{2\hbar}{m_a \omega_a}} \sigma_z (a+a^\dagger) + \sqrt{\frac{2\hbar}{m_b \omega_b}} g_b \sigma_x (b+b^\dagger),
\end{eqnarray}
where $a^\dagger$ and $b^\dagger$ are creation operators for the vibrations for the two degrees of freedom associated with the conical intersection.  

An important question about the quantum dynamics of the chromophore retinal concerns the explanation for the following observation\cite{BenNun02}.  The speed and quantum efficiency of the conformational change is much greater when the retinal is in the rhodopsin environment rather than in a solvent.
This model could be used to address this question by performing the following (non-trivial) calculation. Suppose following optical excitation the TLS is in the excited state and the bath and vibrational modes are in the equilibrium state associated with the ground state of the TLS. We want to calculate the branching ratio between the two possible final states of the system: to the ground electronic state and the excited electronic state and the associated phonon equilibrium state. Transition to the former requires that the non-adiabatic coupling term induce a transition to the ground state by a mechanism similar to Landau-Zener tunneling. However, in the presence of strong decoherence
this tunneling will be suppressed.

We now consider the magnitude and effect  of the coupling of the bath to the TLS near the conical intersection.  In the case of Ohmic dissipation, we can define a dimensionless coupling $\alpha= \eta/2\pi\hbar$.  For the spin-boson model \eqref{eq:spin boson defn} with $\Delta \ll \hbar\omega_c$, if $\alpha>\frac{1}{2}$ and $\epsilon=0$ there will always be incoherent relaxation \cite{Leggett87}, while for $\alpha>1$ and small $\Delta$ the system is localised at its initial state.  For $\epsilon=0$ and $\alpha<\frac{1}{2}$, coherent oscillations between the two states will occur.  These will be inhibited by increasing the bias $\epsilon$ and the temperature \cite{Leggett87} but enhanced by values of $\Delta$ much larger than $\omega_c$ \cite{Carmeli88}.

We can make an estimate for $\alpha$ in water, using the constants given earlier which completely describe the solvent. The two solute unknowns are the cavity size and change in dipole moment.  If we measure these in angstroms and Debye respectively ($1D =3.3\times 10^{-30} C\cdot m$) then $\alpha = 22 \frac{(\Delta\mu)^2}{a^3}$. For a typical chromophore with radius between $5-8\mbox{\AA}$, a dipole moment change of just $1-5D$ is sufficient to make $\alpha>1$, and this condition seems likely to be met for most small molecules \cite[Ch. 6]{Reichardt88}.  In contrast, we note that for GFP, quantum chemistry calculations suggest $\Delta \mu$ is small \cite{Helms02}. The solvent is therefore a large potential source of decoherence which destroys any quantum coherent oscillations of the electronic excitation.

\section{Fluorescence resonant energy transfer}

\begin{figure}[tbp]
\begin{center}
\includegraphics[width=5cm]{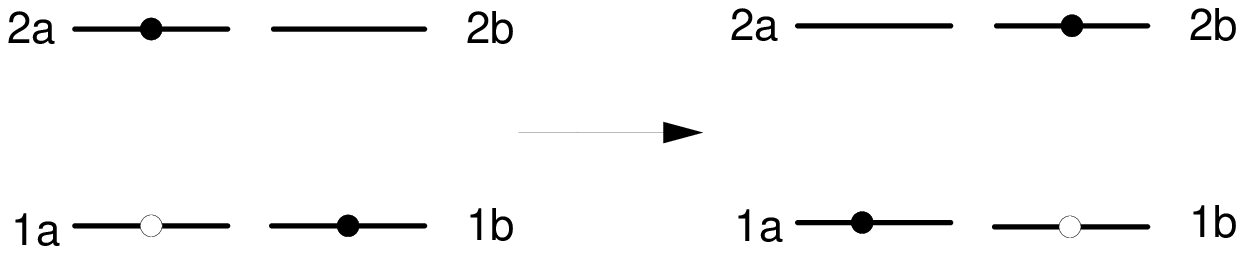}
\caption{Schematic diagram of the non-radiative transfer of an excitation from molecule $a$ to molecule $b$.  Initially, molecule $a$ is in the excited ($2a$) state, while molecule $b$ is in the ground state ($1b$).  When coupled by fluorescence resonant energy transfer, the first molecule is de-excited to the ground state $1a$, while the second molecule transitions to the excited state $2b$.  No photons are exchanged during the process.}
\label{fig:transfer}
\end{center}
\end{figure}

Another important process in biomolecules is coherent transfer of excitons.  An example is fluorescence resonance energy transfer (FRET), as shown in Figure \ref{fig:transfer}.  This occurs in photosynthesis \cite{Hu97PhysicsToday} and is the basis of FRET spectroscopy \cite{Wu94,Ha01} and a proposal for quantum computing based on biomolecules \cite{Lovett03PRB}.  If $1$ and $2$ denote the two molecules, the coupling can be described by adding a term to the Hamiltonian
\begin{eqnarray}
H_0 &= \frac{J_{12}}{2} \left(\sigma_+^1 \sigma_-^2 + \sigma_-^1 \sigma_+^2 \right) = J_{12} \left(\sigma_x^1 \sigma_x^2 + \sigma_y^1 \sigma_y^2 \right),
\end{eqnarray}
where $\sigma^{\pm}=\sigma_x\pm i \sigma_y$.  $J_{12}$ describes the FRET coupling and scales like $1/R^3$ where $R$ is the separation of the molecules \cite{Forster65}.

What about the effect of the solvent? We assume that the molecules are sufficiently far apart that the reaction fields acting on the two biomolecules are uncorrelated.  This assumption can be tested by looking at the wavevector dependence of the dielectric constant $\epsilon(\omega,\vect{q})$; this will be weak.  We anticipate that the corresponding length scale is of the order of the separation of the polar molecules in the solvent which in turn is much larger than $R$, the separation of the biomolecules. Thus, we model the reaction fields by two independent collections of harmonic oscillators.  The relevant Hamiltonian is then
\begin{eqnarray}\label{eq:fret}
H =&&  \frac{1}{2} \epsilon_1\sigma_z^1 + \sum_\alpha \omega_\alpha a_\alpha^\dagger a_\alpha + \sigma_z^1 \sum_\alpha g_\alpha (a_\alpha + a_\alpha^\dagger) \; + \nonumber\\
&& \frac{1}{2} \epsilon_2 \sigma_z^2 + \sum_\beta \omega_\beta b_\beta^\dagger b_\beta \;+ \sigma_z^2 \sum_\beta \widetilde{g}_\beta (b_\beta+b_\beta^\dagger)\; + \nonumber\\
&& J_{12} (\sigma_x^1\sigma_x^2 + \sigma_y^1 \sigma_y^2),
\end{eqnarray}
and $J(\omega),\widetilde{J}(\omega)$ are as before, and $\epsilon_1,\epsilon_2$ are the energy gaps of the two biomolecules. In the Ohmic regime, this will have similarities to two Kondo impurities coupled by an XX interaction and in an external magnetic field \cite{Mahan90b}. We have studied this problem in more detail in a separate article \cite{Gilmore04arXiv}.  There, we show that in most cases the above Hamiltonian may be reduced to a single spin-boson model. This allows us to give quantitative criteria, in terms of experimentally measurable system parameters, that are necessary for coherent Bloch oscillations of excitons between the chromophores, and suggest how these could be observed experimentally through FRET spectroscopy.

The above Hamiltonian \eqref{eq:fret} has a natural generalisation to many coupled biomolecules, such as the rings of chlorophyll molecules in a light harvesting  complex in purple bacteria \cite{Hu97PhysicsToday} or in a segment of base pairs in double stranded DNA \cite{Bouvier02}. It will allow calculation of the localisation (the extent of spatial coherence) of excitons in such extended systems.

\section{Summary}

The major new result of this work is that we have shown that a chromophore and its environment may be modelled by an independent boson model, and derived the appropriate spectral density, which can be expressed in terms of experimentally obtainable parameters.  We have estimated the critical coupling parameter $\alpha$, and found that quantum states are rapidly decohered by the interaction of the chromophore dipole with its surrounding polar environment.  However, as can be seen from equation \eqref{eq:eta}, the coupling to the environment is strongly dependent on the cavity radius, and further work must be done to determine the effects of surrounding proteins and solvation shells which might push the bulk solvent away from the chromophore and reduce the decoherence.  Finally, we have introduced spin-boson models which could be used to model conformational change around a conical intersection and molecule coupled by fluorescence resonance energy transfer.  These models, although simple, exhibit rich physics and may help to provide insight into the workings of these complex systems.

\ack
This work was supported by the Australian Research Council and by the University of Queensland Graduate School Research Travel Awards.  We thank R.H. Austin, G.A.D. Briggs, I. Burghardt,P.C.W. Davies, B.L. Hu, B. Lovett, T. Martinez, P. Meredith,G.J. Milburn, J. Riesz and K. Schulten for helpful discussions.  We thank the QIPIRC at  Oxford for hospitality.

\bibliographystyle{unsrt}
\bibliography{phd}

\begin{thebibliography}{10}
\bibitem{DecayNote}
For example, retinal undergoes non-radiative decay to a specific conformational
  change within 200 fs. In contrast, the excited state of GFP has a lifetime
  that is more than four orders of magnitude longer and undergoes radiative
  decay with a very high quantum efficiency.
\bibitem{DebyeNote}
Note that the sign of the Fourier transform variable used in \cite{Hsu97} is
  opposite to ours, so we have inverted the sign of $\omega$ in \eqref{eq:debye
  model}; otherwise, an incorrect negative sign is introduced.
\end{thebibliography}

\end{document}